# A Computational Model of Children's Learning and Use of Probabilities Across Different Ages

Zilong Wang
Integrated Program in Neuroscience, McGill University

Thomas R. Shultz
Department of Psychology and School of Computer Science, McGill University

Ardavan S. Nobandegani
Department of Psychology and Department of Electrical & Computer Engineering, McGill University

## Abstract

Recent empirical work has shown that human children are adept at learning and reasoning with probabilities. Here, we model a recent experiment investigating the development of school-age children's non-symbolic probability reasoning ability using the Neural Probability Learner and Sampler (NPLS) system. We demonstrate that NPLS can accurately simulate children's probability judgments at different ages, tasks and difficulty levels to discriminate two probabilistic choices through accurate probability learning and sampling. An extension of NPLS using a skewed heuristic distribution can also model children's tendency to wrongly select the outcome with more favorable items but less likely to draw the favorable ones when the probabilistic choices are similar. We discuss the roles of two model parameters that can be adjusted to simulate the probability matching versus probability maximization phenomena in children, and why frequency biases children's probabilistic judgments.

## Introduction

In order to better adapt to an uncertain world, humans and a variety of other species have developed a remarkable ability to learn and reason with probabilities (e.g., Denison et al., 2013; Denison & Xu, 2010, 2014; Eckert et al., 2018; Tecwyn et al., 2017; Xu & Garcia, 2008). This allows them to learn probability distributions and use them to effectively guide their actions (Denison & Xu, 2014; Eckert et al., 2018; Tecwyn et al., 2017).

One prominent phenomenon in this area is probability matching, a decision strategy in which predictions of class membership are proportional to the class base rates. The ubiquity of probability matching is well supported by experimental studies across many species, e.g., bees (Greggers & Menzel, 1993), fish (Behrend & Bitterman, 1961), turtles (Kirk & Bitterman, 1965), human infants and adults (Denison & Xu, 2014; Eckert et al., 2018), and apes (De Petrillo & Rosati, 2019; Eckert et al., 2018). Relatedly, a neural network model called *Neural Probability Learner and Sampler* (NPLS) has been used to simulate a series of empirical studies of probability matching in human infants (Shultz & Nobandegani, 2021).

A recent empirical study demonstrated that school-aged children also exhibited probability matching and sometimes used raw frequency when making their choices in probabilistic discrimination tasks (O'Grady & Xu, 2020). Here, we model this study of school-aged children, thereby complementing previous simulations of human adults and infants, and explaining the use of both probability matching and frequency by children in probabilistic tasks.

## Experiments

O'Grady and Xu (2020) designed two experiments which required children to choose one collection of marbles out of two (in the form of pictures of marbles on two trays) that is more likely to yield a marble of a target color on a single random draw, i.e., the collection that has a larger proportion of the target marbles. For example, one collection has 70% target marbles and 30% non-target marbles whereas the other collection contains 20% target marbles and 80% non-target marbles. Clearly, the former "winning" collection should be chosen instead of the latter "losing" collection.

In Experiment 1 there were 24 six- and 24 seven-year-old children and two conditions: *total equal* and *target equal*. In the *total equal* condition both collections had 100 marbles, whereas in the *target equal* condition both collections contained the same number of target color marbles with the losing collection containing more nontarget marbles and therefore a smaller proportion of target marbles. Importantly, the proportions of target marbles in the two collections varied across trials, therefore yielding different ratios of proportions, referred to as RoR[1]. RoR is calculated as:

$$\frac{\textit{Proportion of target marbles in the winning collection}}{\textit{Proportion of target marbles in the losing collection}}$$

---

[1] RoR stands for Ratio of Ratios, which is used interchangeably for the Ratio of Proportion of target marbles.

where the numerator is always larger than the denominator.

Human ability to form abstract approximations of numerical magnitude is often explained by use of the Approximate Number System (ANS) (Dehaene & Changeux, 1993) which follows Weber's Law and shows ratio dependence between two magnitudes (Halberda & Feigenson, 2008; Pica et al., 2004; Whalen et al., 1999, as cited in (O'Grady & Xu, 2020)). The ANS is assumed to play a key role in discriminating between two proportions and thus RoR is used as a metric here for probability discrimination. RoR varied from 1.1 to 14 such that proportions were relatively closer and harder to discriminate for smaller RoRs but further apart and easier for larger RoRs.

In Experiment 2 there were 40 eight-year-old, 50 ten-year-old, and 40 twelve-year-old children, and two conditions, *total equal* and *frequency vs probability*. Like Experiment 1, in the *total equal* condition, each pair of collections had the same total number of marbles, which varied across trials. In the *frequency vs probability* condition, the losing collection contained more target marbles than the winning collection, but with enough nontarget marbles to make its proportion of target marbles smaller than the winning collection's. RoR varied from 1.1 to 9.5.

In both experiments, accuracy was recorded and statistically analyzed, showing that accuracy increased with RoR in a logarithmic shape and that the overall acuity of probability estimations increased with age. At each age, easier trials, with larger RoR, yielded higher accuracy than more difficult trials, with smaller RoR.

## Methods

We adopt the NPLS system (Shultz & Nobandegani, 2020, 2022) used to simulate the chimpanzee and human adults experiments of Eckert et al. (2018) and human infant experiments of Denison and Xu (2014).

NPLS has two parts: learning and sampling. The learning part is an adapted version of the sibling-descendant cascade-correlation (SDCC) neural-network learning algorithm that has been successful in modeling many deterministic phenomena in cognitive and language development (Shultz, 2003, 2017). SDCC is a constructive algorithm that builds a network by recruiting hidden units as needed (Baluja & Fahlman, 1994). The sampling part is a Markov Chain Monte Carlo (MCMC) sampling algorithm that has been used to simulate a wide range of empirical findings in human probabilistic reasoning and decision making (Dasgupta et al., 2017; Nobandegani & Shultz, 2020). By sampling from the learned probability distribution, MCMC mimics an agent's selection process.

Like many learning neural networks, SDCC networks learn from examples by reducing overall prediction error. They are deterministic, feed-forward networks (Baluja & Fahlman, 1994) in which unit activations are passed forward from input units that depict examples to hidden units that transform inputs into more abstract representations, and finally to output units coding the target response to those particular inputs. Network output can be considered an expectation of what will happen, while target output represents what actually happens.

SDCC processing has an input phase and an output phase. During learning (output phase), network error $E$ is reduced by adjusting connection weights:

$$E = \sum_o \sum_p (A_{op} - T_{op})^2 \quad (1)$$

where $A$ is the actual output activation for unit $o$ and pattern $p$, and $T$ is the target output activation for this unit and pattern.

SDCC training starts with only the input and the output layer, and then adds hidden units one at a time as needed to solve the problem being trained. Thus, the SDCC algorithm builds its own network topology without being hand designed. In the input phase, input weights to candidate hidden units are trained to maximize the covariation of each candidate hidden unit's output activation with network error $E$. Then the highest correlating unit is installed either on the highest layer of hidden units or on its own higher layer, depending on which has a higher covariation with the current network error (Baluja & Fahlman, 1994). Input weights are frozen once the recruited units are installed so that, when the network switches to output phase, it adjusts weights only one layer at a time, never needing to (unrealistically) back-propagate error signals. SDCC maximizes the covariance $C$ between candidate-hidden-unit activation and overall network error in input phases:

$$C = \frac{\sum_o |\sum_p (h_p - \langle h \rangle)(e_{op} - \langle e_o \rangle)|}{\sum_o \sum_p (e_{op} - \langle e_o \rangle)^2} \quad (2)$$

where $h_p$ is activation of the candidate hidden unit for pattern $p$, $\langle h \rangle$ is the mean activation of the candidate hidden unit for all patterns, $e_{op}$ is the residual error at output $o$ for pattern $p$, and $\langle e_o \rangle$ is the mean residual error at output $o$ for all the training patterns.

For the task of probability learning, SDCC networks use an asymmetric sigmoid activation function:

$$y_i = \frac{1}{1 + e^{-x_i}} \quad (3)$$

where $y$ is the output activation of receiving unit $i$, $x$ is the net input to unit $i$, and $e$ is the exponential function. Output activation thus ranges from 0 to 1, resembling probabilities.

Because of the deterministic nature of SDCC, it would recruit new hidden units infinitely on probabilistic problems. To avoid this, NPLS monitors its error reduction progress over the learning iterations. SDCC is already equipped to monitor the progress within both input and output phases using parameters for threshold and patience. In output phase, error is reduced by adjusting connection weights. When error reduction stagnates, the algorithm shifts to input phase to recruit a new hidden unit, adjusting weights entering candidate units to increase the correlation between their activations and network error. In both phases, stagnation is detected when there is no progress greater than a threshold parameter for the number of training epochs specified by a patience parameter. This scheme is extended by adding another loop with its own threshold and patience parameters to monitor advancement over learning iterations, where each iteration contains an input phase and the next output phase (Shultz & Doty, 2014). This learning cessation mechanism allows NPLS to halt when error reduction stagnates over iterations, which turns out to accurately match the ground-truth probability. Thus, NPLS can learn any unnormalized multivariate probability distribution from examples specifying whether or not an output occurs in the presence of a particular input (Kharratzadeh & Shultz, 2016).

We run the same number of NPLS networks as the number of children in each age group in the respective RoR and experimental conditions. Networks are trained on event sequences with the input unit arbitrarily coding for the identity of the marble collection (1 or 2) and an output unit with 1 representing a target marble and 0 representing a non-target marble, thus directly corresponding to the visual stimuli the children experienced. In this way, networks learn to output the probability of drawing a target marble from each collection of marbles. Because ground-truth probabilities are not used as learning targets, probability estimates represented by learned network output activations are an emergent property of NPLS learning.

A simplified example of this coding is shown in Table 1, comparing the two frequency ratios of 2:1 and 3:4. Input values of 1 and 2 distinguish the two collections. Output values depict the presence (1) or absence (0) of a target marble. In this example of a 2:1 frequency ratio, there are 2 instances of the target marbles (each coded as 1) and 1 instance of non-target marble (coded as 0). Likewise, outputs for the 3:4 frequency ratio contain 3 target (each coded as 1) and 4 non-target (each coded as 0) marbles. More abstractly, we can describe the two output codes in equations involving frequency ratios $a:b$ and $c:d$, where, in this case, $a = 2$, $b = 1$, $c = 3$, and $d = 4$.

Table 1. Example coding of frequency ratios

| 2:1 | | 3:4 | |
| --- | --- | --- | --- |
| Input | Output | Input | Output |
| 1 | 1 | 2 | 1 |
| 1 | 1 | 2 | 1 |
| 1 | 0 | 2 | 1 |
| | | 2 | 0 |
| | | 2 | 0 |
| | | 2 | 0 |
| | | 2 | 0 |
| Mean | 2/3 = .67 | Mean | 3/7 = .43 |

The ground-truth probabilities for drawing the target marble can then be expressed as $a / (a + b)$ for input 1 and $c / (c + d)$ for input 2. Expressed verbally, 2 out of 3 marbles in collection 1 are of the target color, while 3 out of 7 marbles in collection 2 are of the target color.

This example problem presents a particularly interesting comparison as the raw frequency of the preferred marble color is higher in collection 2 than in collection 1 (3 > 2), while the probability of randomly drawing the preferred-color marble is higher in collection 1 than in collection 2 (.67 > .43). Such problems allow us to distinguish between selection strategies of frequency and probability in both humans and computer agents (Shultz & Nobandegani, 2022).

In the *frequency vs probability* condition, the empirical data suggested that 8- to 12-year-old children used a selection strategy based on frequency for smaller RoRs. To simulate such a strategy, we use a closely analogous coding scheme. Consistent with our model for comparing frequency ratios, we extend SDCC to a broader and more generalizable psychological model for number comparison (Shultz et al., 2022) for comparing integer values. The key difference is that this comparison focuses only on the *a* and *c* values representing the raw frequencies of target marble, completely ignoring the *b* and *d* values that represent non-target marbles. In this case, we can regard the output activations for the two frequencies being compared as belief strengths in picking the larger one, rather than matching ground-truth probabilities. Thus, the two selection strategies yield different answers: the frequency strategy selects collection 2 (3 > 2), while the probability strategy selects collection 1 (.67 > .43).

A parameter called score-threshold is used to distinguish deeper learning in older children from shallower learning in younger children. Formally, score-threshold indicates the allowance between actual and expected outputs that are considered as correct. Smaller score-thresholds yield more accurate outputs. Assuming that learning is deeper for older children, we set score-threshold to .53 for 6-year-old group and .52 for 7-year-old group in Experiment 1; .51 for 8-year-old group, .50 for 10-year-old group and .49 for 12-year-old group in Experiment 2.

With recent advances in how to probabilistically generate examples from learned categories (Nobandegani & Shultz, 2017), NPLS then uses a Markov Chain Monte Carlo (MCMC) algorithm to simulate how participants select examples mentally, essentially transforming a deterministic neural network into a probabilistic generative network. NPLS learns categories from examples and then generates examples from those categories using the learned weights (Nobandegani & Shultz, 2018). Such backward inferences can be mathematically characterized as a form of sampling from the underlying, learned probability distribution. Analogously, a child could mentally draw a sample, cued by the higher probability of a target object, and thus identifying the more favorable collection for obtaining that object.

Formally, NPLS induces a probability distribution $p(\mathbf{X}|\mathbf{Y})$ on the deterministic input-output mapping $f(\mathbf{X}; W^*)$ learned by an NPLS network and uses MCMC to sample from that induced distribution. The induced distribution is given by:

$$p(\mathbf{X}|\mathbf{Y} = Y) \propto exp(-\beta ||Y - f(\mathbf{X}; W^*)||_2^2) \qquad (4)$$

where $||\cdot||_2$ is the l2-norm, $W^*$ the set of weights for a network after training, and $\beta$ a damping factor. For an input instance $\mathbf{X} = X$ belonging to the desired class $Y$, the network output $f(X; W^*)$ is expected to be close to $Y$ in the $l_2$-norm sense. Equation 4 adjusts the probability of input instance $X$ to be inversely proportional to the base-e exponentiation of the $l_2$ distance. The NPLS system can handle any MCMC method, including Metropolis-Adjusted Langevin, a gradient-based MCMC method, which could be implemented in a biologically-plausible way (Moreno-Bote et al., 2011; Savin & Denève, 2014).

To estimate probability matching in the context of RoR, it is tempting to assume that the "probability" that is matched with is proportion of target marbles in either collection. However, the two proportions often do not sum to 1, thus can't be considered as probability. Instead, "probability" should ideally be matched on either proportion divided by the sum of them. Here we only consider ideally matching the probability of choosing the winning collections correctly.

In order to integrate probability and frequency strategies in our simulation for the *frequency vs probability* condition, we introduce a linear probability distribution (Figure 1) such that there is a higher chance of choosing by frequency when RoR is smaller and a higher chance of choosing by probability when RoR is larger. Because the older age groups increase faster in accuracy, their slope is steeper than the younger group.

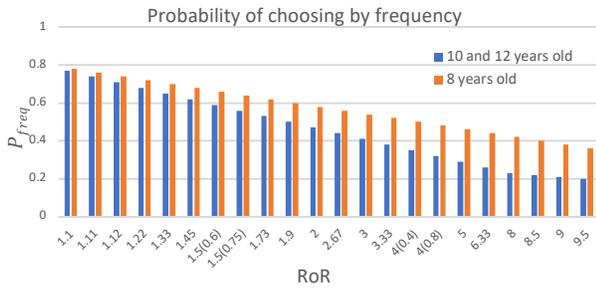

Figure 1: $P_{freq} = 1 - P_{prob}$ where $P_{freq}$ represents the chance of choosing by frequency and $P_{prob}$ represents the chance of choosing by probability.

## Results

We plot the mean network probability estimates, which are the mean output activations of the networks, with standard deviations, and the ground-truth proportions of target marbles in the winning and losing collections, for each RoR in each age group in Experiment 1 (Figures 2 and 3) and Experiment 2 (Figures 4 and 5). Simulated agents representing older children are more accurate and have less variance than those for younger children. Mean network output activations correlate highly with ground-truth probabilities across all conditions in both experiments. Pearson *r* correlations between ground-truth probabilities and mean output activations are presented under the corresponding legends.

**Ground Truth Probabilities and Mean Network Probability Estimates**

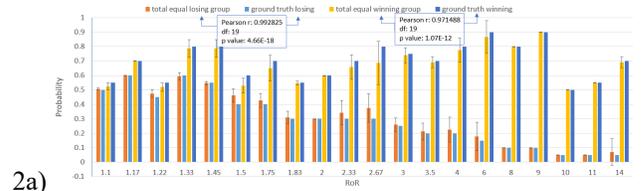

2a)

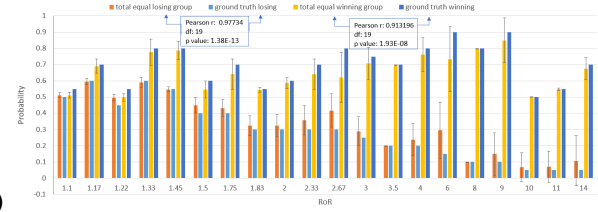

2b)

Figure 2: Experiment 1 Total Equal: score-threshold = 0.52, 7-year-old (2a); 0.53, 6-year-old (2b).

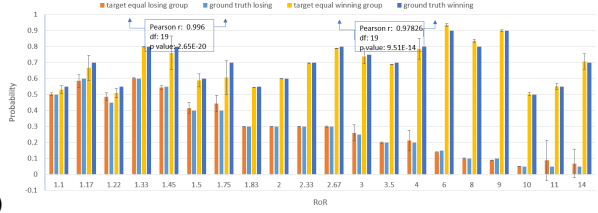

3a)

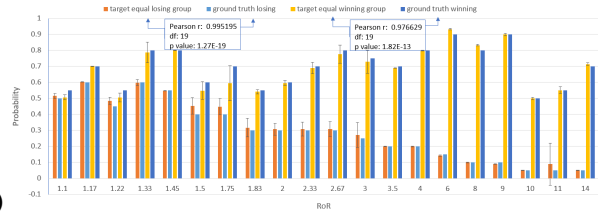

3b)

Figure 3: Experiment 1 Target Equal: score-threshold = 0.52, 7-year-old (3a); 0.53, 6-year-old (3b).

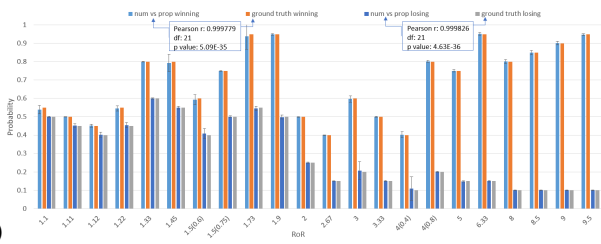

4a)

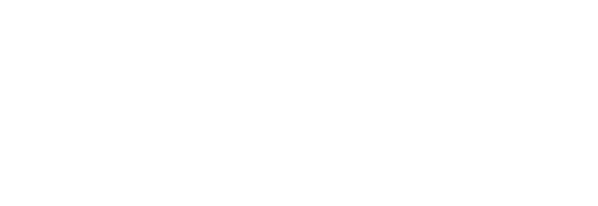

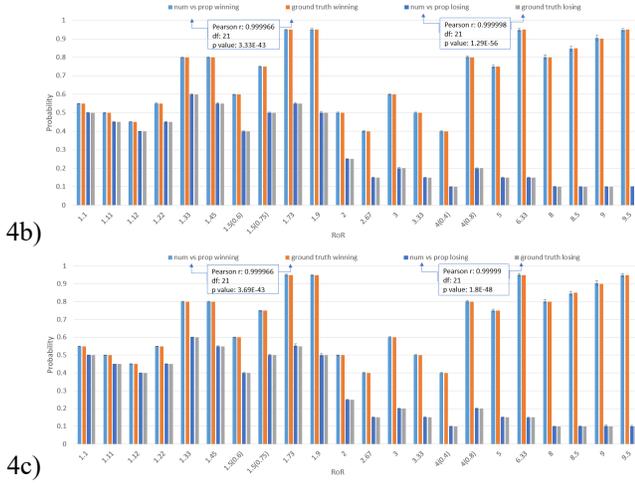

4b)

4c)

Figure 4: Experiment 2 frequency vs probability: score-threshold = 0.51, 8-year-old (4a); 0.5, 10-year-old (4b); 0.49, 12-year-old (4c).

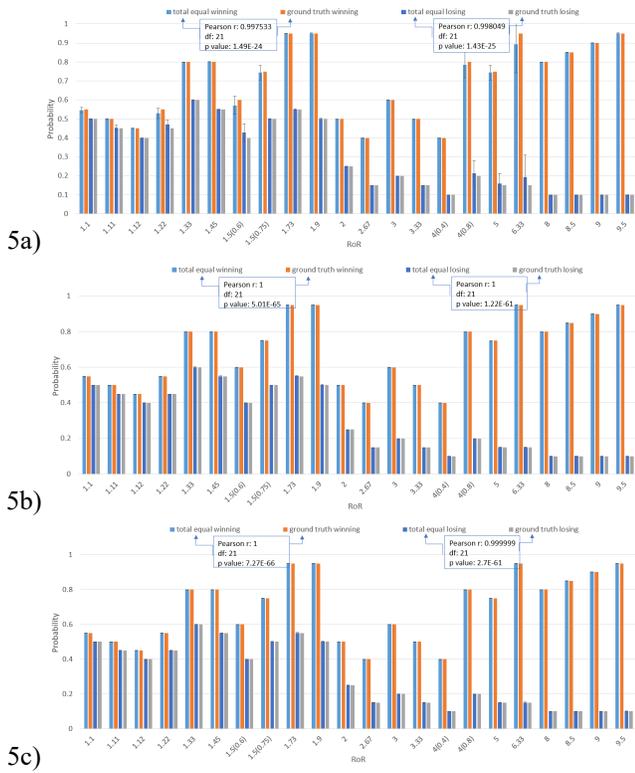

5a)

5b)

5c)

Figure 5: Experiment 2 Total Equal: score-threshold = 0.51, 8-year-old (5a); 0.5, 10-year-old (5b); 0.49, 12-year-old (5c).

## MCMC Mean Sampling Probability

We also plot the MCMC sampling means with standard deviations for each age group, task, and experiment. For each plot, we present the ideal sampling accuracy under probability matching and probability maximization and specify the correlation between sampling means and ideal samping accuracy under probability matching. Each sampling result corresponds to the empirical sampling data (Figure 10). For the total-equal conditions in both Experiment 1 (Figure 6) and 2 (Figure 9) MCMC sampling successfully captures the logarithmic trend shown in children's data (Figure 10). The NPLS agents in Experiment 1 choose correctly with a probability around 0.5 when RoR is as low as 1.1, and gradually reach an accuracy of 1 for the older group and 0.8 for the young group at the highest RoR value of 14. Similarly, the NPLS agents in Experiment 2 start at 0.5 and 0.6 accuracies for the young and older groups, and eventually increase to around 0.9 and 1 respectively at the highest RoR value of 9.5. Sampling also resembles the trend in children's target equal condition (Figure 7), with NPLS agents choosing correctly around 50% of the time at a lower RoR of 1.1 and reaching an accuracy of 0.8 to 1 for the older group and 0.6 for the younger group at the highest RoR of 14.

Notably, the sampling result for *frequency vs probability* condition (Figure 8) correlates highly with the empirical data (for 12-year-old group, $r(20) = 0.9362$, 95% CI = [0.8502, 0.9735], $p < 10^{-9}$; for 10-year-old group, $r(20) = 0.9380$, 95% CI = [0.8542, 0.9743], $p < 10^{-9}$; for 8-year-old group, $r(20) = 0.8960$, 95% CI = [0.7625, 0.9564], $p < 10^{-7}$). For the youngest age group, agents choose correctly constantly below chance for RoR lower than 2. For example, they only have an accuracy of 0.2 when RoR is 1.1, and they choose more correctly as RoR increases. The probability of choosing correctly reaches 0.6 for the highest RoR of 9.5. NPLS agents representing the other two older age groups are more accurate with respect to higher RoR and have a probability of choosing correctly of 0.8 at RoR of 9.5. Although they still have an accuracy lower than 0.5 when RoR is below some value smaller than 2, it is better than the youngest group. It is worth noting that although the Pearson correlations with ideal probability matching are extremely high, this doesn't indicate probability matching because the sampling accuracies are not equivalent.

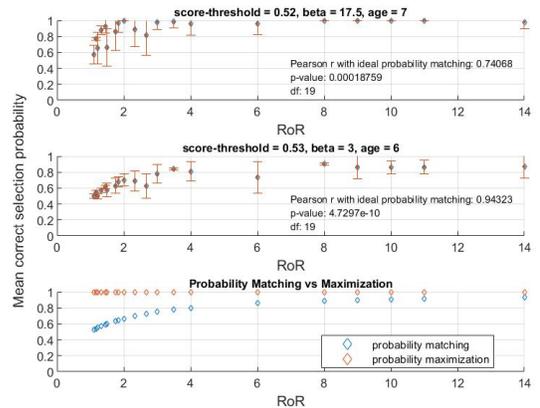

Figure 6: Experiment 1 Total Equal.

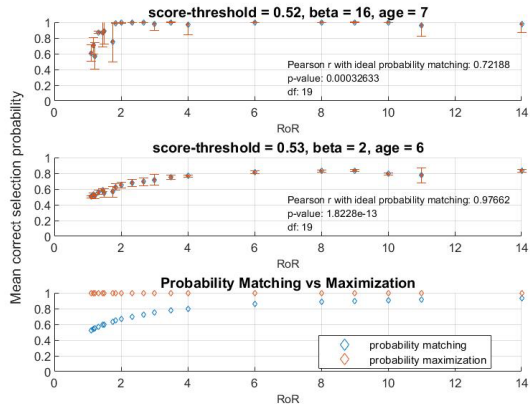

Figure 7: Experiment 1 Target Equal.

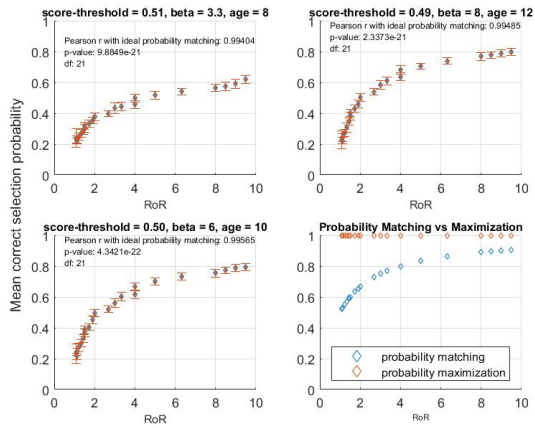

Figure 8: Experiment 2 Frequency vs Probability.

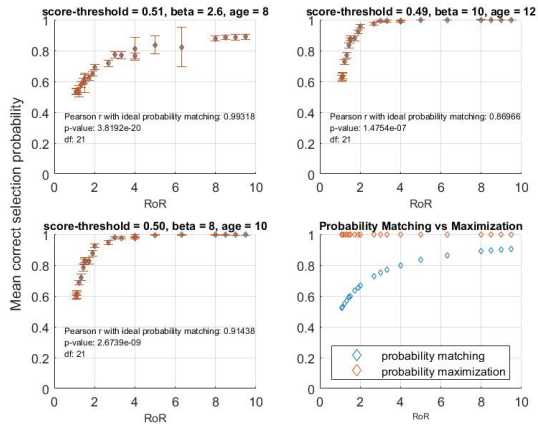

Figure 9: Experiment 2 Total Equal.

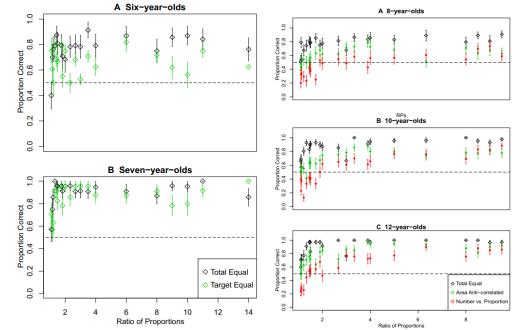

Figure 10: Empirical Data (O'Grady & Xu, 2020)

We also conduct a linear regression analysis (Table 2) to show that the mean probability for sampling the winning collection increases linearly with log (RoR), equivalent to increasing logarithmically with raw RoR.

Table 2: Linear regression of log (RoR) with sampling probability for each task and age.

| Expt | Task | Age | Β | p < | $R^2_{adj}$ |
|---|---|---|---|---|---|
| 1 | Total Equal | 6 | 0.5466 | 0.0001 | 0.822 |
| | | 7 | 0.7742 | 0.0008 | 0.444 |
| | Target Equal | 6 | 0.5311 | 0.0001 | 0.880 |
| | | 7 | 0.7823 | 0.0018 | 0.395 |
| 2 | Frequency vs Probability | 8 | 0.1712 | 0.0001 | 0.986 |
| | | 10 | 0.2610 | 0.0001 | 0.967 |
| | | 12 | 0.2585 | 0.0001 | 0.958 |
| | Total equal | 8 | 0.5363 | 0.0001 | 0.963 |
| | | 10 | 0.7056 | 0.0001 | 0.731 |
| | | 12 | 0.7491 | 0.0001 | 0.645 |

## Discussion

Our simulation results show that NPLS accurately models the empirical pattern of children's learning and use of probability across ages seen in O'Grady and Xu's (2020) experiments. Although O'Grady and Xu only dealt with selection and not learning, we assume that participants would have to register the probabilities in some way before using them to guide their actions. Such registration is here modeled in NPLS learning. This is similar to the Eckert et al. (2018) experiment that is also simulated with NPLS (Shultz & Nobandegani, 2022).

Two parameters are particularly important in this simulation: the score-threshold in SDCC and the β damping factor in MCMC. Although accuracy decreases as score-thresholds increase, all score thresholds from 0.53 to 0.49 yield fairly to perfectly accurate learning of the probability. Moreover, β modulates the extent of probability matching. Increase in β moves sampling from probability matching to maximization smoothly. Different βs account for the difference in children's sampling behaviors at different ages. We see a higher Pearson correlation with probability matching in younger children and harder tasks but more probability maximization in easier tasks and older groups. As they get older, children begin to rationally select the winning collection of marbles as much as possible. The reason may be that children concentrate on exploring the probability space before exploiting it when facing difficult choices.

As O'Grady and Xu (2020) explained their empirical data with ANS that supposedly follows Weber's Law, both the empirical data and our simulation show the Weber's effect reported in Eckert et al. (2018) and Shultz & Nobandegani's (2022) studies, that is, correct sampling probability increases logarithmically with raw RoR. However, the improvement of accuracy varies greatly across different tasks and age groups, which suggests that ANS is not sufficient to account for all the empirical data. ANS at best yields frequency information, which is ignored in favor of probabilities, when the two sources of information are disentangled. Our simulation of *frequency vs probability* provides an alternative explanation.

When probability and frequency are disentangled in *frequency vs probability* conditions, 8- to 12-year-old children systematically chose the wrong collection for the lower RoR, presumably because they were using the frequency heuristic. However, previous work has shown that infants use probability, not frequency (Denison & Xu, 2014), as simulated by NPLS (Shultz & Nobandegani, 2020). This change of behaviour at older ages might be explained by evidence that teaching children about whole numbers biases their understanding of probabilities towards whole numbers (Ni & Zhou, 2005). Our model suggests that children have a changing probability distribution of choosing the frequency heuristic, with a larger likelihood of using the frequency heuristic when RoR is small and decreases as RoR increases. This suggests that a whole-number bias could temporarily interfere with using probabilities, especially when children are faced with highly similar probabilistic choices. However, the overall trend of sampling is still highly consistent with probability matching, which further proves the ubiquity of probability matching even with the interference of frequency.

## References

Baluja, S., & Fahlman, S. E. (1994). *Reducing Network Depth in the Cascade- Correlation Learning Architecture*. CMU-CS-94-209, 1–11. http://www.dtic.mil/docs/citations/ADA289352

Behrend, E. R., & Bitterman, M. E. (1961). Probability-Matching in the Fish. *The American Journal of Psychology*, 74(4), 542. https://doi.org/10.2307/1419664

Dasgupta, I., Schulz, E., & Gershman, S. (2017). Where do hypotheses come from? *Cognitive Psychology*, 96, 1–25.

De Petrillo, F., & Rosati, A. G. (2019). Rhesus macaques use probabilities to predict future events. *Evolution and Human Behavior*, 40(5), 436–446.

Dehaene, S., & Changeux, J. P. (1993). Development of elementary numerical abilities: A neuronal model. *Journal of Cognitive Neuroscience*, 5(4), 390–407. https://doi.org/10.1162/jocn.1993.5.4.390

Denison, S., Reed, C., & Xu, F. (2013). The emergence of probabilistic reasoning in very young infants: Evidence from 4.5- and 6-month-olds. In *Developmental Psychology* (Vol. 49, pp. 243–249). American Psychological Association. https://doi.org/10.1037/a0028278

Denison, S., & Xu, F. (2010). *Twelve- to 14-month-old infants can predict single-event probability with large set sizes*. 1–6. https://doi.org/10.1111/j.1467-7687.2009.00943.x

Denison, S., & Xu, F. (2014). The origins of probabilistic inference in human infants. *Cognition*, 130(3), 335–347.

Eckert, J., Call, J., Hermes, J., Herrmann, E., & Rakoczy, H. (2018). Intuitive statistical inferences in chimpanzees and humans follow Weber's law. *Cognition*, 180, 99–107.

Greggers, U., & Menzel, R. (1993). Memory dynamics and foraging strategies of honeybees. *Behavioral Ecology and Sociobiology*, 32(1), 17–29. https://doi.org/10.1007/BF00172219

Kharratzadeh, M., & Shultz, T. (2016). Neural implementation of probabilistic models of cognition. *Cognitive Systems Research*, 40, 99–113.

Kirk, K., & Bitterman, M. E. (1965). Probability-learning by the turtle. *Science*, 148(3676), 1484–1485. https://doi.org/10.1126/science.148.3676.1484

Moreno-Bote, R., Knill, D. C., & Pouget, A. (2011). Bayesian sampling in visual perception. *Proceedings of the National Academy of Sciences of the United States of America*, 108(30), 12491–12496.

Ni, Y., & Zhou, Y. Di. (2005). Teaching and learning fraction and rational numbers: The origins and implications of whole number bias. *Educational Psychologist*, 40(1), 27–52. https://doi.org/10.1207/s15326985ep4001_3

Nobandegani, A., & Shultz, T. (2018). Example generation under constraints using cascade correlation neural nets. *Proceedings of the 40th Annual Meeting of the Cognitive Science Society*, 2385–2390. http://mindmodeling.org/cogsci2018/papers/0456/0456.pdf

Nobandegani, A., & Shultz, T. (2020). A Resource-Rational, Process-Level Account of the St. Petersburg Paradox. *Topics in Cognitive Science*, 12, 417–432. https://doi.org/6

Nobandegani, A., & Shultz, T. (2017). Converting cascade-correlation neural nets into probabilistic generative models. In G. Gunzelmann, A. Howes, T. Tenbrink, & E. J. Davelaar (Eds.), *Proceedings of the 39th Annual Conference of the Cognitive Science Society* (pp. 1029–1034). Cognitive Science Society.

O'Grady, S., & Xu, F. (2020). The Development of Nonsymbolic Probability Judgments in Children. In

*Child Development* (Vol. 91, Issue 3, pp. 784–798). https://doi.org/10.1111/cdev.13222

Savin, C., & Denève, S. (2014). Spatio-temporal representations of uncertainty in spiking neural networks. In Z. Ghahramani, M. Welling, C. Cortes, N. D. Lawrence, & K. Q. Weinberger (Eds.), *Advances in Neural Information Processing Systems 27* (pp. 2024–2032). Curran Associates, Inc.

Shultz, T. (2003). *Computational developmental psychology*. MIT Press.

Shultz, T. (2017). Constructive artificial neural-network models for cognitive development. In N. Budwig, E. Turiel, & P. D. Zelazo (Eds.), *New Perspectives on Human Development* (pp. 13–26). Cambridge University Press.

Shultz, T., & Nobandegani, A. (2022). Modeling the Learning and Use of Probability Distributions in Chimpanzees and Humans. *Proceedings of the Annual Meeting of the Cognitive Science Society*, 1947–1953. https://escholarship.org/uc/item/41z847j4

Shultz, T., & Nobandegani, A. (2020). Probability without counting and dividing: a fresh computational perspective. In S Denison, M. Mack, Y. Xu, & B. Armstrong (Eds.), *Proceedings of the 42nd Annual Conference of the Cognitive Science Society* (pp. 1–7). Cognitive Science Society.

Shultz, T., Nobandegani, A., & Wang, Z. (2022). A Neural Model of Number Comparison with Surprisingly Robust Generalization. *ArXiv:2210.07392*. https://doi.org/https://doi.org/10.48550/arXiv.2210.07392

Tecwyn, E. C., Denison, S., Messer, E. J. E., & Buchsbaum, D. (2017). Intuitive probabilistic inference in capuchin monkeys. *Animal Cognition*, *20*(2), 243–256. https://doi.org/10.1007/s10071-016-1043-9

Xu, F., & Garcia, V. (2008). Intuitive statistics by 8-month-old infants. *Proceedings of the National Academy of Sciences of the United States of America*, *105*(13), 5012–5015. https://doi.org/10.1073/pnas.0704450105